\documentclass[pre,twocolumn,superscriptaddress,showpacs]{revtex4}
\usepackage{graphicx}
\usepackage{amsmath}
\usepackage{color}
\begin{document}

\title{Influence of long-range interactions on charge ordering phenomena on a square lattice}
\author{Louk Rademaker}
\email{rademaker@lorentz.leidenuniv.nl}
\affiliation{Institute-Lorentz for Theoretical Physics, Leiden University, PO Box 9506, NL-2300 RA Leiden, The Netherlands}
\author{Yohanes Pramudya}
\affiliation{National High Magnetic Field Laboratory, Florida State University, 1800 E. Paul Dirac drive, Tallahassee, FL 32310, USA}
\author{Jan Zaanen}
\affiliation{Institute-Lorentz for Theoretical Physics, Leiden University, PO Box 9506, NL-2300 RA Leiden, The Netherlands}
\author{Vladimir Dobrosavljevi\'{c}}
\affiliation{National High Magnetic Field Laboratory, Florida State University, 1800 E. Paul Dirac drive, Tallahassee, FL 32310, USA}
\date{\today}

\begin{abstract} 
Usually complex charge ordering phenomena arise due to competing interactions. We have studied how such ordered patterns emerge from the frustration of a long-ranged interaction on a lattice. Using the lattice gas model on a square lattice with fixed particle density, we have identified several interesting phases; such as a generalization of Wigner crystals at low particle densities and stripe phases at densities in between $\rho = 1/3$ and $\rho = 1/2$. These stripes act as domain walls in the checkerboard phase present at half-filling. The phases are characterised at zero temperatures using numerical simulations, and mean field theory is used to construct a finite temperature phase diagram.
\end{abstract}


\maketitle

\section{Introduction}

The formation of ordered structures is one of the main topics in the field of condensed matter physics. Starting from the relative straightforward crystalline order a wide variety of increasingly complex ordering phenomena has been observed and proposed such as stripes\cite{Zaanen:1989p1602,Tranquada:1995p5405,Tranquada:1994p5659,Cheong2000,Salamon:2001p5657,Dagotto:2005p5310,Boothroyd:2011p5350,Andrade:2012p5422}, charge density waves\cite{GRUNER:1988p5325}, incommensurate phases\cite{BAK:1982p5293} and so forth.

These complex ordering patterns usually arise as a result of competing interactions. For example, the kinetic energy of holes competes with the tendency towards antiferromagnetic order in cuprates thus forming stripes. In ANNNI models, the next-nearest neighbor Ising coupling has the opposite sign as the nearest neighbor coupling. The question immediately arises whether higher-order commensurate or incommensurate phases can appear in systems with only one type of interaction.

Of course this is the case. In the continuum the sole presence of long-range interactions will cause particles to form a Wigner crystal. When a fixed number of particles are placed on an underlying lattice the desired Wigner crystalline order may be incommensurate with the lattice, thus leading to frustration.

\begin{figure}
 \includegraphics[width=\columnwidth]{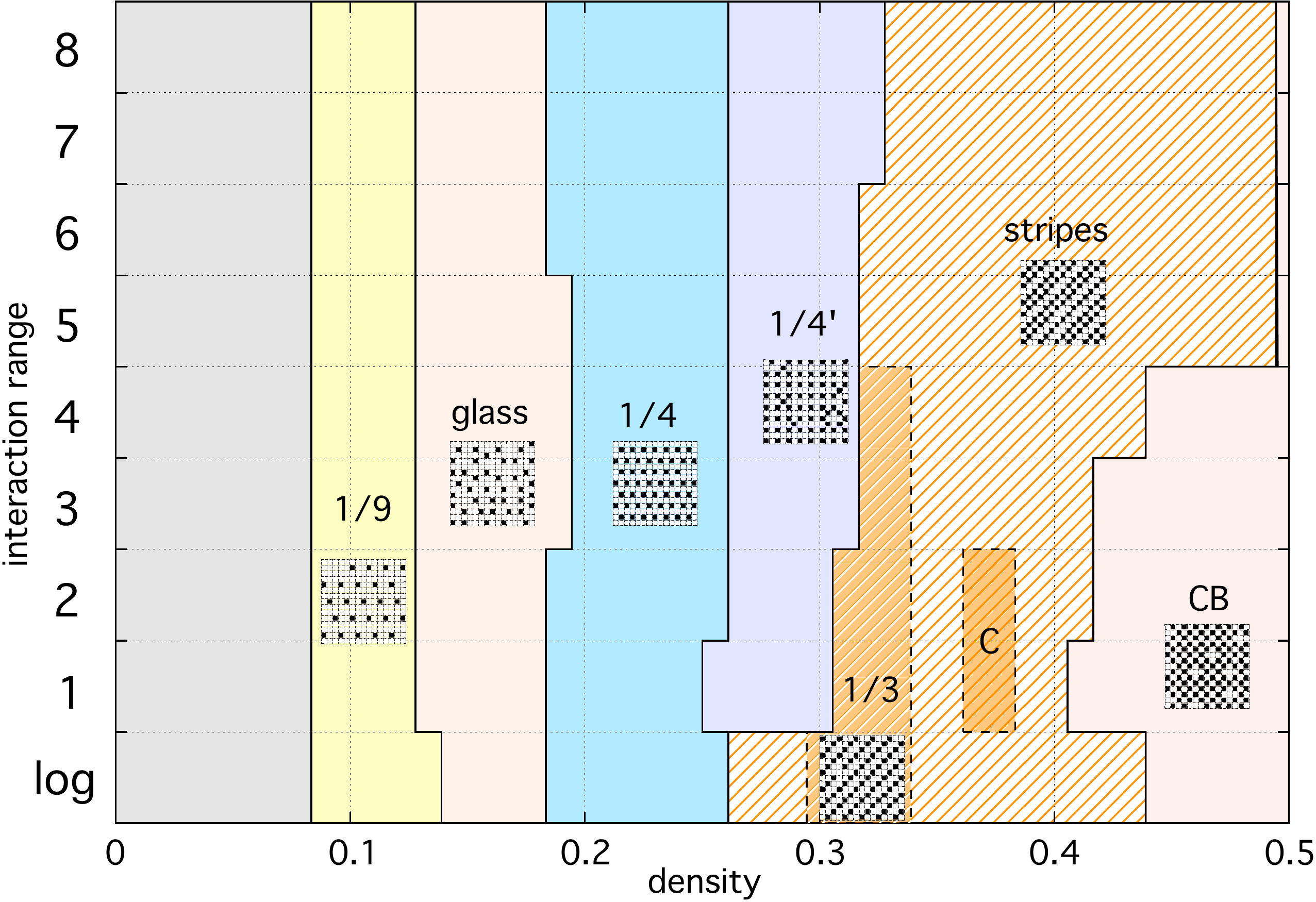}
 \includegraphics[width=\columnwidth]{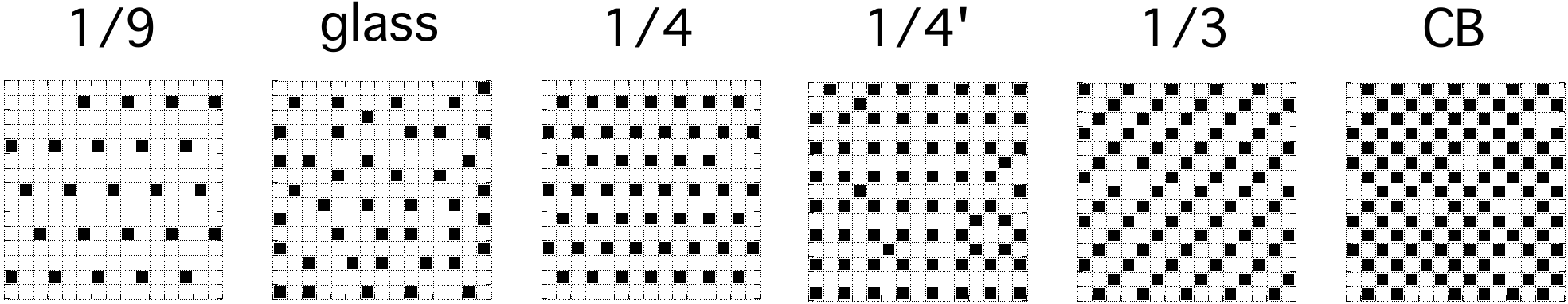}
 \caption{\label{PD}(Color online) The approximate ground state phase diagram of the long-range lattice gas model on a square lattice, based on variational methods as discussed in section \ref{SectionNumerics}. On the vertical axis the type $p$ of the long-range interaction $V(r) = \frac{1}{r^p}$ is given, together with a logarithmic decaying interaction. The horizontal axis represents the particle density. From low to high densities we identify the following phases: The area without name depicts the dilute generalized Wigner crystal, followed by the $1/9$ Wigner crystal, the $1/6$ glassy phase described by \cite{Lee:2002p5305}, the $1/4$ Wigner crystal, the `checkerboard-in-a-checkerboard' $1/4$' phase, stripe phases (with a plateau for the $1/3$ stripe phase and `C' denotes the channelled stripes as described by \cite{Lee:2001p5312}) and finally the checkerboard phase. Phases below $1/4$ filling are discussed in section \ref{SectionWigner}, above $1/4$ filling are discussed in section \ref{SectionStripes}. Below the phase diagram typical particle configurations in six phases are shown enlarged.}
\end{figure}

\begin{figure}
 \includegraphics[width=\columnwidth]{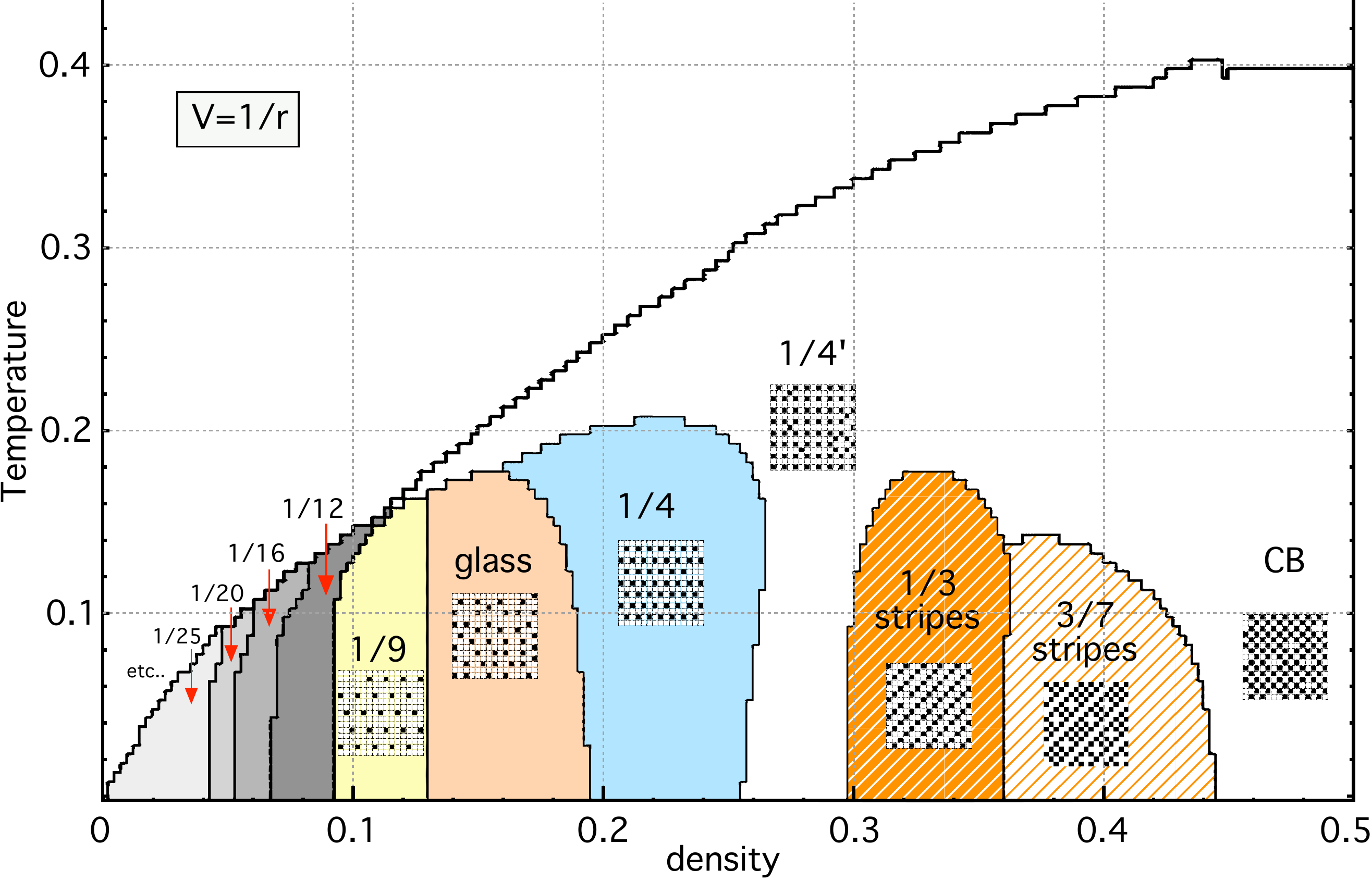}
 \caption{\label{PDfiniteT}(Color online) Mean field finite temperature phase diagram of the lattice gas model (\ref{LGModel}) on a square lattice with $V \sim 1/r$ interactions, see section \ref{SectionFiniteT}. Temperature is in units of the nearest neighbor interaction. The phases are the same as in the zero-temperature phase diagram of Figure \ref{PD}. At low densities we find various Wigner crystalline phases (see section \ref{SectionWigner}) with densities of the form $1/pq$ with $p,q$ integers. Close to half-filling we find checkerboard order which has a smooth crossover to the `checkerboard-in-a-checkerboard' $1/4'$ phase. Around $\overline{n}=1/3$ and $3/7$ there are stripe ordered phases (see section \ref{SectionStripes}). The transitions towards the $1/4'$ and checkerboard phase are second order, the other transitions are first order.}
\end{figure}

We have investigated the influence of long-range interactions on charge ordering phenomena on a square lattice. Expanding the results of Ref. \cite{Lee:2001p5312,Lee:2002p5305} we explored the full range of particle densities $0 \leq \rho \leq 1$ and types of long-range interactions $V = 1/r^p$. Our main result is summarized in Figures \ref{PD} (zero temperature) and \ref{PDfiniteT} (finite temperature), where we depict phase diagrams of unusual charge ordered patterns. At low densities the competition between the continuum triangular Wigner lattice and the underlying square lattice indeed leads to a plethora of `generalised Wigner' crystals. At higher densities, this leas to variations of the checkerboard pattern which is well-known at half-filling. Stripe phases appear as they are rooted in the topological defects of the checkerboard order.

We do not claim that the phase diagrams we derived are the exact phase diagrams. As is often the case for frustrated systems, a large set of metastable states persists down to zero temperature. Unbiased numerical computation of the energy for a large ensemble of configurations gives us a strong indication that indeed the phase diagram of Figure \ref{PD} is correct, however, these indeed be metastable states incorrectly recognized as ground states.

The lay-out of this paper is as follows. In section \ref{SectionLGM} we introduce the lattice gas model, which is the model describing interacting classical particles on a lattice. In the two subsequent sections we discuss qualitatively the ordered structures at low densities (section \ref{SectionWigner}) and at densities close to half-filling (section \ref{SectionStripes}). We have performed a Monte Carlo simulation in section \ref{SectionNumerics} to derive the zero temperature phase diagram of Figure \ref{PD}. In section \ref{SectionFiniteT} we extend these results to finite temperatures using mean field theory, see Figure \ref{PDfiniteT}.

\section{Long-range lattice gas models}
\label{SectionLGM}

The lattice gas model can be defined on any kind of lattice but we focus only on the square lattice. On each of the $N$ lattice sites there can be a particle or not, denoted by $n_i = 1$ or $0$ respectively. These particles interact via some general potential $V_{ij}$. The corresponding Hamiltonian is then
\begin{equation}
	H_L = \sum_{i \neq j} V_{ij} (n_i-\rho) (n_j-\rho) - \mu \sum_{i} n_i.
	\label{LGModel}
\end{equation}
We subtract the average particle density $\rho$ to prevent divergent energies. In the grand-canonical ensemble, the chemical potential $\mu$ tunes the average particle density $\rho N = \sum_{i} n_i$. The model (\ref{LGModel}) is in fact equivalent to the Ising model\cite{Lee:1952p5483}. Under the replacement $\sigma^z_i = 2 n_i - 1$ and considering only a nonzero nearest neighbor interaction $\frac{1}{4} V_{\langle ij \rangle} = J$ one finds
\begin{equation}
	H_I = J \sum_{\langle ij \rangle} \sigma^z_i \sigma^z_j - B \sum_{i} \sigma^z_i.
\end{equation}
The chemical potential maps onto an external magnetic field $B = \frac{1}{2} \mu - V_{\langle ij \rangle}$, while the particle density maps onto the mean magnetization.

For the ferromagnetic Ising model the ground state is completely magnetically ordered, which amounts to either a full or empty lattice in the lattice gas parlance. In addition, a model with antiferromagnetic coupling will be half-filled with particles if the external magnetic field is small, $|B| < 2 J$. Therefore, using the standard grand canonical ensemble will in general not enable us to investigate all possible particle densities: a canonical ensemble - fixed particle number - is required. 

We argue that most physical realizations of lattice gas models are in fact at a fixed particle number, and not at fixed chemical potential. One particular example is the oxygen ordering in YBCO planes, where it is beyond doubt that the number of oxygen ions in the lattice is fixed\cite{DEFONTAINE:1990p5300}. The patterns in which the oxygen ions align themselves are quasi-one-dimensional, in a manner similar to the expected electronic ordering in TTF-TCNQ salts \cite{Hubbard:1978p5308}. Whilst studying the latter, Hubbard has developed a general solution for the ground state of a lattice gas model with long-range interactions at any particle density in one dimension. Hubbard's solution only requires the interaction energy as a function of distance to be convex.

Among two-dimensional realizations of lattice gas models are for example the ordering of ad-atoms on a surface\cite{Pokrovsky:1978p5320,BAK:1982p5293,Feng:2011p5317}, XY systems\cite{Villain:1977p5322}, higher order commensurate magnetic phases\cite{Fisher:1980p5294,BAK:1982p5293,SELKE:1988p5281} or stripe order in high-temperature superconductors\cite{Low:1994p5313,Emery:1993p5321,Zhang:2003p5301}. Especially systems with anisotropic short-ranged interactions or competing short- and long-range interactions\cite{Giuliani:2011p5684,Giuliani:2013p5685} have acquired considerable attention over the years, therefore we wish to focus here to the case of long-range isotropic interactions\cite{Lee:1992p5307,Mobius:2009p5314,Troster:2010p5303,Pramudya:2011p5302}.

Most studies of lattice gas models in two dimensions however restrict their attention to half-filled, empty and full lattices, due to the aforementioned grand-canonical reasons. There are two notable exceptions: the stripe order discussed in Ref. \cite{Lee:2001p5312} between $1/3$ and $1/2$ filling and the glassy dynamics at $1/6$ filling\cite{Lee:2002p5305}. These results were obtained for a `quasi-logarithmic' repulsive interaction which is a solution of Poisson's equation on a lattice, $\nabla^2 V_{ij} = - 2 \pi \delta_{ij}$. Given the nontrivial ordering patterns discovered there, as a follow-up we present here a systematic study of the ground state orderings at all densities between $0$ and $1/2$, for general repulsive interactions
\begin{equation}
	V_{ij} = \frac{1}{|r_{ij}|^p} >0.
	\label{InteractionModel}
\end{equation}
In the next two sections we will first discuss qualitatively such long-range lattice gas models at fixed densities, while in the remaining sections the picture will be further quantified using numerical simulations and mean field theory.

\section{Dilute densities - Generalized Wigner crystals}
\label{SectionWigner}

\begin{figure}
 \includegraphics[width=0.5\columnwidth]{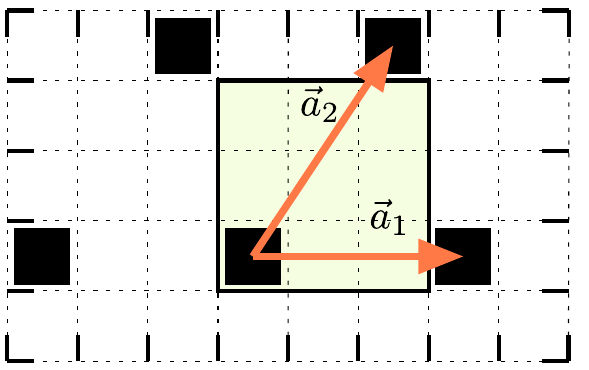}
 \caption{\label{Fig1_9}(Color online) At density $\rho=1/9$ a triangular crystal of particles is formed, which is not equilateral as would be the case for a perfect Wigner crystal. It is thus a prime example of a generalized Wigner crystal. The unit cell and unit vectors of the Wigner crystal are shown.}
\end{figure}

In the previous section we introduced the lattice gas model, that we will now study at fixed densities on the square lattice with long-range repulsive interaction of the form (\ref{InteractionModel}). In the limit of very low particle density, the underlying square lattice becomes irrelevant compared to the average inter-particle distance,
\begin{equation}
	\ell_p \gg a
\end{equation}
where $a$ is the lattice constant. In the continuum description of particles repelled by a long-range force, Wigner\cite{Wigner:1934p5323} showed that the interaction energy is minimized when the particles form a crystalline structure, which is triangular in two dimensions\footnote{Wigner\cite{Wigner:1934p5323} considered quantum particles in second order perturbation theory with $1/r$ interactions and compared the energies of various different lattice structures. In his work the triangular lattice had the lowest energy. Even though no other structure has been found with a lower energy, the Wigner crystallization into a triangular lattice is not rigorously proven. For classical particles studied in this publication the exact energy is equal to the 'quantum' second order perturbation result since that amounts to the expectation value of the interaction energy. For logarithmic interactions we refer to studies of vortex lattices that indicate that triangular lattices are in that case the lowest energy configurations\cite{Kleiner:1964p5694}. For $p>1$ interactions we compared the energy of the triangular lattice with the square lattice, which again favors the triangular lattice.}. In general, one can state that the energy of a Wigner crystalline state of the particles is\cite{Pokrovsky:1978p5320}
\begin{equation}
	E = J N \rho \sum_{\mathrm{crystal}} \frac{1}{|d|^p}
\end{equation}
where $N$ is the number of lattice sites of the underlying lattice, $\rho$ is the particle density and the summation runs over the particles in the Wigner crystal. The distance between particles in the Wigner crystal $\ell_p$ scales with the inverse square root of the density. Therefore the energy in the low density limit scales as
\begin{equation}
	E \propto J N \rho^{p/2 + 1}.
\end{equation}
The presence of the underlying lattice is now a source of frustration. Ref. \cite{Pokrovsky:1978p5320} considers an underlying triangular lattice, leading to frustration only if the density is not of the form $1/p^2$. For example, when $\rho=1/9$ a perfect triangular Wigner crystal can be formed. On top of a square lattice, however, it is not possible to form a triangular Wigner crystal because $\sqrt{3}$ is irrational. One can nevertheless construct `almost perfect' triangular crystals. As an example, consider the density $\rho = 1/9$, see Figure \ref{Fig1_9}. The lowest energy state is there also a triangular crystal of particles, but not equilateral as for a perfect Wigner crystal. In principle such `almost perfect' Wigner crystals could exist at densities
\begin{equation}
	\rho_{pq} = \frac{1}{pq}
	\label{CrystalD}
\end{equation}
with $p, q$ integers while $p \leq q$, such that the following equilateral triangle relation is approximated by
\begin{equation}
	p \sim \frac{1}{2} \sqrt{3} q.
\end{equation}
For example, the densities $\rho = 1/9$, $1/12$, $1/16$ etcetera would allow such `almost perfect' triangular crystal. Following the work of Hubbard\cite{Hubbard:1978p5308} in one dimensional systems, we will call these particle orderings `generalized Wigner crystals'. From this qualitative reasoning we argue that such Wigner crystals \emph{might exist}. Note, however, that one has to resort to a numerical computation to find whether such crystals have indeed the lowest energy at a given density. 

\begin{figure} 
  \includegraphics[width=\columnwidth]{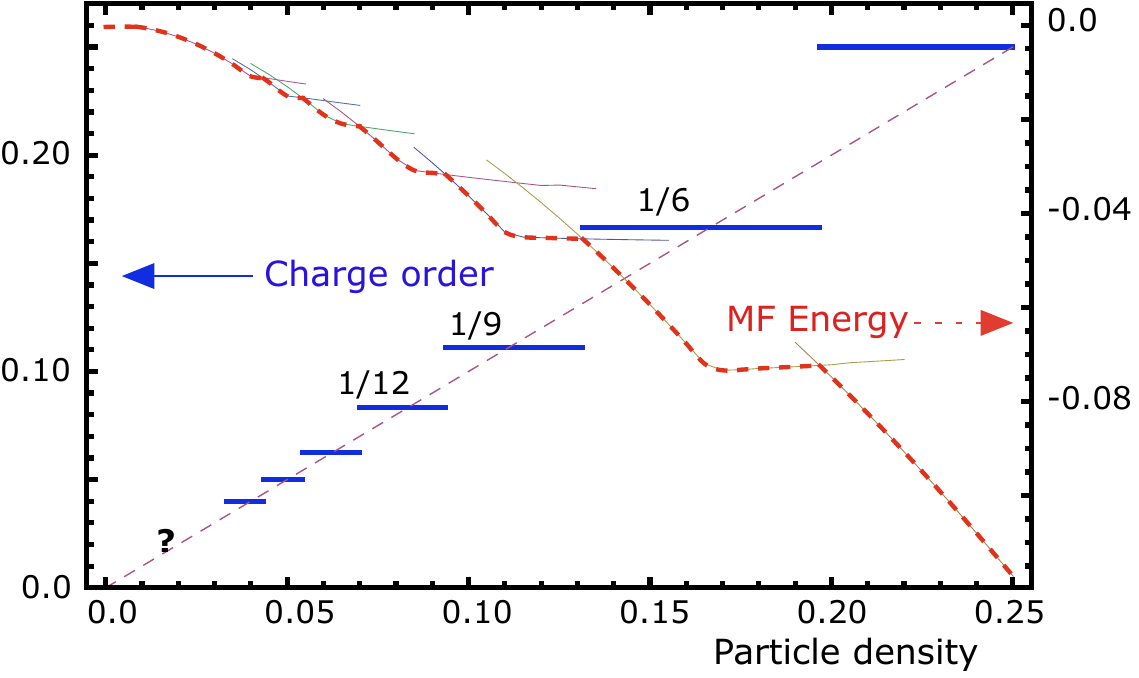}
  \caption{(Color online) A generalized Wigner crystal can be classified according to peaks in the Fourier transformed particle density. For densities close to each specific crystal density (\ref{CrystalD}) the associated crystal structure will be maintained. We put forward the hypothesis that this leads to a devil's staircase of generalized Wigner crystals at low densities. The figure shows the crystal structure versus density obtained by mean field theory (see section \ref{SectionFiniteT}) at $\beta \rightarrow \infty$, displaying a staircase (thick solid lines). The thermodynamic potential (\ref{ThermoPot}) relative to the disordered state is shown (decreasing red dashed line), the thin lines underneath indicate the energy of specific ordered states.}
  \label{DevilsFig}
\end{figure}

So far we only considered densities of the form $\rho_{pq} = 1/pq$. When the density of a lattice gas is in between such densities, we suspect that it is favorable to maintain a generalized Wigner crystal structure. The deviation from the $\rho_{pq}$ density can be accommodated by a superlattices of crystal defects or interstitial vacancies. If such a superlattice of defects forms, still the original $\rho_{pq}$ order is visible in for example the Fourier transformed particle density, where each crystal type has its own specific Fourier peaks. We expect therefore a 'plateau' at densities in the vicinity of each specific $\rho_{pq}$, where the associated crystal structure remains intact modulo the interstitial superlattice. As we will describe in more detail later when discussing the numerical results, indeed a 'plateau' is observed for the $1/4$, $1/6$ and $1/9$ states. Using mean field theory, we went as far as the $1/25$ crystal phase, as is shown in Figure \ref{DevilsFig}. We have therefore strong indications the plateau-structure exists all the way to $\rho \rightarrow 0$, yielding an infinite staircase of plateaux. This structure is reminiscent of the \emph{devil's staircase}, as exists in the case of the one-dimensional lattice gas in the grand-canonical ensemble\cite{Bak:1982p5683}, where specific charge orderings are stable for a finite window of chemical potential. 

Notice that starting at $1/4$ filling the generalized Wigner crystal picture certainly fails. If one adds one single particle to the $1/4$ crystal, it will be necessarily next to another particle. Since the nearest-neighbor repulsion is the strongest, and nearest-neighbor occupancy can be avoided for any density below half-filling, the $1/4$ crystal will be quickly destroyed upon adding particles. For densities above $1/4$ it is necessary to start reasoning from the ordering occurring at the half-filled lattice.

\section{Domain walls and stripes}
\label{SectionStripes}

Exactly at half-filling the ground state is `checkerboard' like, or antiferromagnetic in the Ising language. This means that one sublattice is exactly filled and the other sublattice is completely empty. Densities slightly less than half-filling can be obtained by removing particles from the checkerboard pattern, a process we call hole doping. The density of holes $\rho_h$ is defined as follows
\begin{equation}
	\rho_h = 1 - 2\rho
\end{equation}
where $\rho$ is the total particle density. The same scaling arguments for the dilute particle limit $\rho \ll 1$ can be applied to the dilute hole limit $\rho_h \ll 1$, so close to half-filling the energy scales as
\begin{equation}
		E \propto E_{1/2} \left( 1 - 2 \rho_h \right) + JN c_h \rho_h^{p/2+1}
		\label{ScaleHF}
\end{equation}
where $E_{1/2}$ is the energy of the half-filled checkerboard configuration and $c_h$ is some proportionality constant.

\begin{figure}
	a. \includegraphics[height=1.6cm]{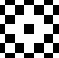}
	b. \includegraphics[height=1.6cm]{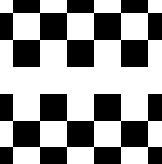}
	c. \includegraphics[height=1.6cm]{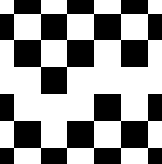}
	d. \includegraphics[height=1.6cm]{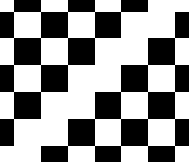}
	\caption{\label{FigDomainwalls} Some examples of domain walls between checkerboard phases. \textbf{a.} A particle on the 'wrong' sublattice surrounded by 'holes', which forms the smallest possible domain wall loop. \textbf{b.} A horizontal domain wall. \textbf{c.} Instability of a horizontal domain wall, by moving one particle to the other domain. \textbf{d.} Stable, diagonal domain walls can exist for all long-range interactions. }
\end{figure}

This scaling argument assumes that all particles will remain on the filled sublattice of the checkerboard phase. However, the checkerboard phase is a broken sublattice symmetry phase and therefore domain walls and topological defects can exist between regions where the checkerboard phase is realized on different sublattices. Though the ground state at densities slightly less than half-filled might be unrelated to the checkerboard phase, we discuss in this section possible structures that are related to the checkerboard. Thus the smallest example, which is neglected in the scaling arguments of equation (\ref{ScaleHF}), is shown in Figure \ref{FigDomainwalls}a. There a single particle on the `wrong' sublattice is surrounded by holes, which is obviously a stable configuration.

On a large scale domain walls may exist such as the one depicted in Figure \ref{FigDomainwalls}b. However, such a straight domain wall is not stable. One can imagine a single particle moving to the other side, thus causing the domain wall to meander. The energy difference between the configurations in Figure \ref{FigDomainwalls}b and Figure \ref{FigDomainwalls}c is given by the energy of that single moved particle,
\begin{eqnarray}
	\Delta E &=& E_{\mathrm{straight}} - E_{\mathrm{meander}}
	\nonumber \\ &	\sim &
		\sideset{}{'}{\sum}_{n \; \mathrm{even}} 
		\left( \frac{1}{|n|^p} - \frac{1}{(n^2+1)^{p/2}} \right) >0
\end{eqnarray}
where the prime on the summation means that we should exclude $n=0$. Since the meandering domain wall has a lower energy than the straight one, the latter is unstable. This argument can be pursued further to the point where one finds that only a diagonal domain wall, as shown in Figure \ref{FigDomainwalls}d, is locally stable. The resulting domain wall has a surface energy that vanishes in the thermodynamic limit; but is extremely stiff with respect to bending.

\begin{figure}
	a. \includegraphics[height=2.4cm]{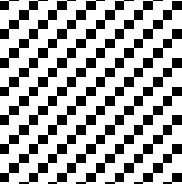}
	b. \includegraphics[height=2.4cm]{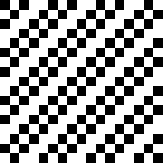}
	c. \includegraphics[height=2.4cm]{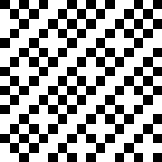}
	\caption{\label{StripeF} Some examples of stripe phases. \textbf{a.} The $\rho = 1/3$ state. \textbf{b.} The $\rho = 2/5$ state. \textbf{c.} The $\rho = 3/7$ state, but now doped.}
\end{figure}

Now a single domain wall on a infinite lattice will not affect the average particle density. However, to obtain particle densities away from half-filling one can introduce a macroscopic number of domain walls. Such a periodic array of domain walls constitutes a `stripe phase', similar to the ones discussed in cuprates\cite{Zaanen:1989p1602,Tranquada:1995p5405,Dagotto:2005p5310}, nickelates\cite{Tranquada:1994p5659}, manganites\cite{Cheong2000,Salamon:2001p5657,Dagotto:2005p5310} or cobaltates\cite{Boothroyd:2011p5350,Andrade:2012p5422}. Examples can be seen in Figure \ref{StripeF}.

If the ground state of a long-range lattice gas model is perfectly stripy, then the system is effectively reduced to a one-dimensional system. The arguments of Hubbard\cite{Hubbard:1978p5308} then apply and one can thus find the specific stripe ordering, as is shown in Figure \ref{StripeF}a and b.
 
But again, for some densities it may pay off not to form perfect stripes but rather to dope a stripe structure as in Figure \ref{StripeF}c. It is then a matter of numerical computation to find out whether the ground state is a Hubbard-type stripe pattern or a `doped' stripe pattern. Lee, Lee and Kim\cite{Lee:2001p5312} call these doped stripe patterns `partially filled diagonal channels'. Finally, we must emphasize that the discussion in this section does in no way whatsoever constitute a proof of existence of stripe ordered or hole doped checkerboard phases. The only way to find the state with the lowest energy is a tedious numerical computation.

\section{Simulations and characterisation of the phases}
\label{SectionNumerics}

Let us now describe the numerical algorithm that was used to find the lowest energy configurations. First notice that for each form of order there is symmetry breaking and hence a degeneracy in the ground state. As the simplest example, observe that the checkerboard configuration at half-filling is two-fold degenerate. However, for the purpose of finding the specific type of charge order, we do not need to worry about ground state degeneracy.

In our algorithm we took various different types of initial configurations, such as the generalized Wigner crystals of section \ref{SectionWigner}, stripe structures of section \ref{SectionStripes}, Wigner crystal structures of defects in the checkerboard, variations of the checkerboard phase, suggestions from literature and a large ensemble of random configurations. We then swapped filled and empty sites randomly. A swap is accepted if it lowers the total energy of the system. The long-range nature of the interaction was taken into account by summing the interaction over all mirror charges as in an Ewald summation\cite{Essmann:1995p5324}. For the quasi-logarithmic interaction we followed the method of Ref. \cite{Lee:2001p5312}.

The major issue is that one cannot know for sure whether this algorithm leads to the global ground state or that one gets stuck in a local energy minimum. Indeed, for a frustrated system we expect to find a large number of metastable states. The method of simulated annealing, by which we mean slowly reducing the temperature to zero, was therefore also used to avoid getting stuck in a local energy minimum.  Upon comparing the energies of configurations obtained from the various initial configurations, using both zero temperature swapping and simulated annealing, we found a lowest energy state at each density.

Our work was performed on a square lattice with $90 \times 90$ and $154 \times 154$ sites. These lattice sizes were chosen such that the linear dimension $L$ is divisible by the prime numbers $2, 3, 5$ (for $L=90$) and $7$ and $11$ (for $L=154$). We looked at all densities that were a multiple of the linear dimension, hence $\rho = n/90$ and $\rho=n/154$ for all integers $n=1 \ldots L/2$.

\begin{figure}
	\includegraphics[width=\columnwidth]{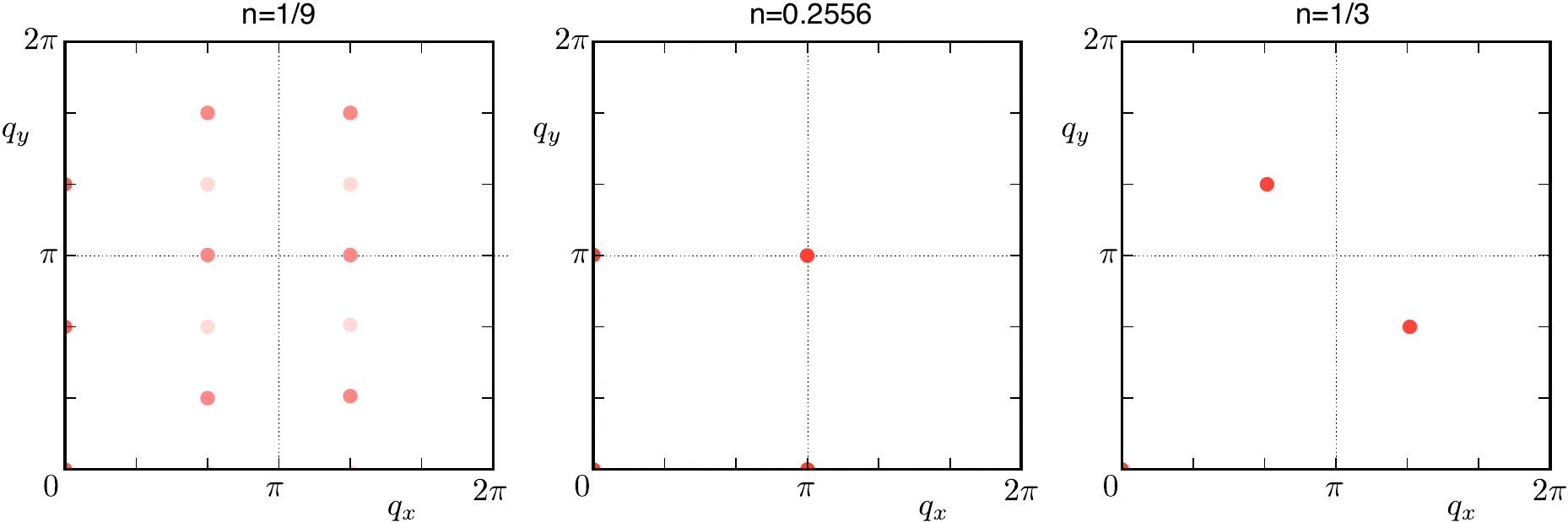}
	\caption{\label{FigFourier}(Color online) Fourier transformed density at fillings $1/9$, $0.25556$ and $1/3$; on a $90 \times 90$ lattice with $1/r$ interactions. The different ordering wave vectors are clearly visible. These peaks are used to identify the phases that lead to the phase diagram of figure \ref{PD}.}
\end{figure}

\begin{figure}
	\includegraphics[width=\columnwidth]{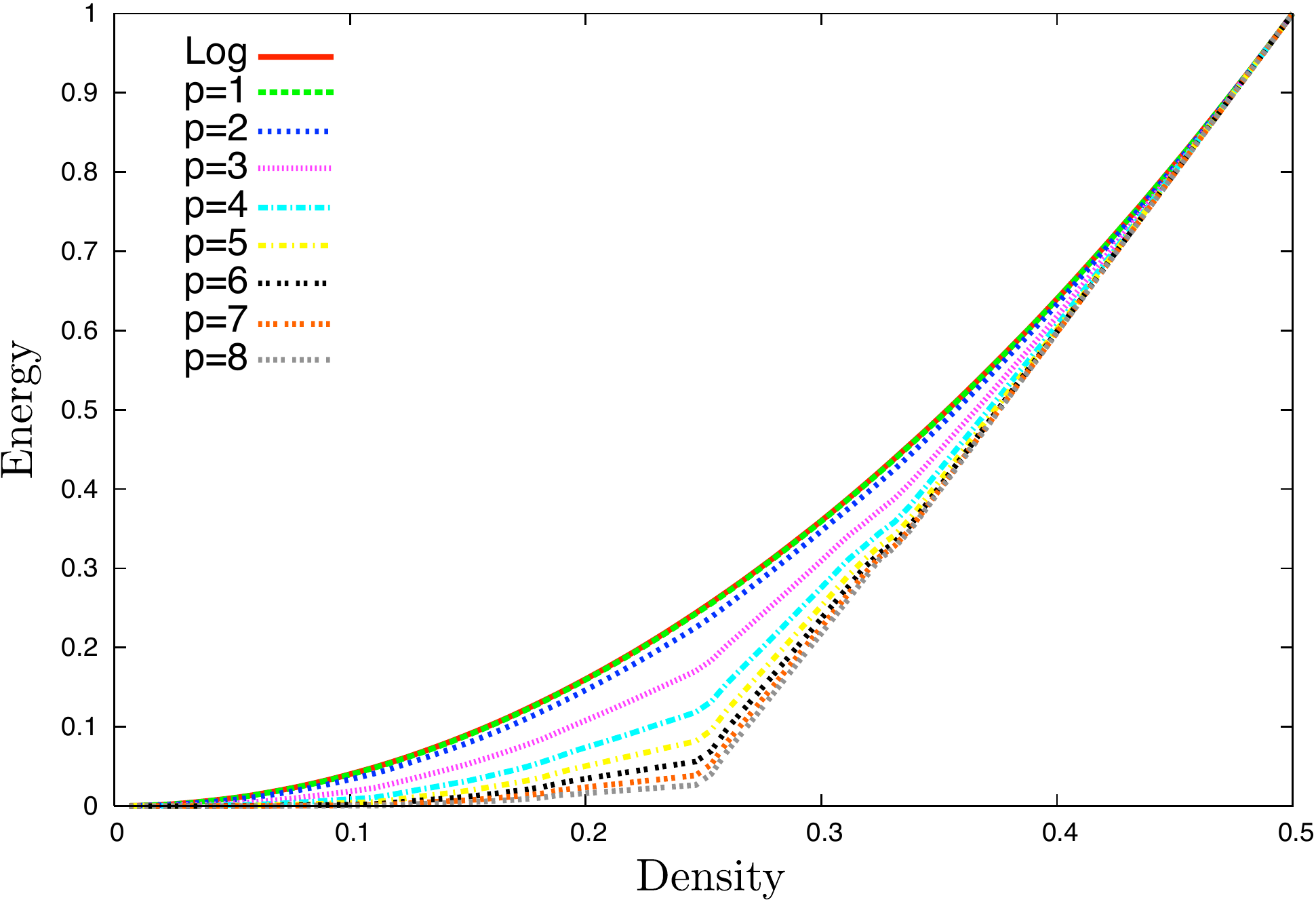}
	\caption{\label{Energy}(Color online) Ground state energy of the long-range lattice gas model as a function of density, for various types of interaction. All energies are rescaled such that $E(\rho=1/2) = 1$. Starting with a logarithmic interaction in solid red, we computed the energy for interactions of the form $1/r^p$. Notice the scaling behavior in the limit $p\rightarrow \infty$ and notice the kink around $\rho = 1/3$ signaling stripe order.}
\end{figure}

After obtaining the ground state configuration, a Fourier transform of the particle density was taken,
\begin{equation}
	n(k) = \frac{1}{N} \sum_i e^{-ikr_i} n_i.
\end{equation}
The peaks in the Fourier spectrum were used to identify the specific orderings at each density, as can be seen in Figure \ref{FigFourier}. The ground state energy as a function of density is shown in Figure \ref{Energy}, rescaled such that the energy at half-filling equals $E=1$. Indeed the general scaling behavior close to zero and half-filling, as described in the previous two sections, is retrieved. This is most explicit in the limit of $p \rightarrow \infty$, the energy becomes constant between $\rho = 0$ and $\rho = 1/4$, and linear between $\rho=1/4$ and $\rho=1/2$. Notice an extra kink in the energy around $\rho=1/3$, which signals the onset of the stripe order.

With the caution that the resulting configurations might be in fact metastable states incredibly close to the actual ground state, we constructed the zero temperature phase diagram in Figure \ref{PD}. At low densities the finite lattice size used in our numerical simulations form a limitation with regard to the precision of the results. The first unequivocal observed charge ordering state is the $1/9$ generalized Wigner crystal phase, which is stable for a considerable range of densities around $\rho = 1/9$. Interestingly, the presence of this phase is remarkably independent of the interaction range $p$.

The $1/9$ phase is followed by the $1/6$ phase which is extensively discussed in the work of Ref. \cite{Lee:2002p5305}. There the $1/6$ phase is characterized as a glassy phase, with infinite ground state degeneracy due to the infinite ways one can tile unit cells of $2 \times 3$ lattice sites. For further details we refer to Ref. \cite{Lee:2002p5305}.

Directly below $\rho = 1/4$ densities the $1/4$ generalized Wigner crystal phase is stable. As described before, introduction of particles to densities higher than $\rho=1/4$ leads to a superlattice of interstitials which can be called a `checkerboard-in-a-checkerboard $1/4'$ phase', which seems to be absent in the case of logarithmic interactions.

\begin{figure}
	\includegraphics[width=\columnwidth]{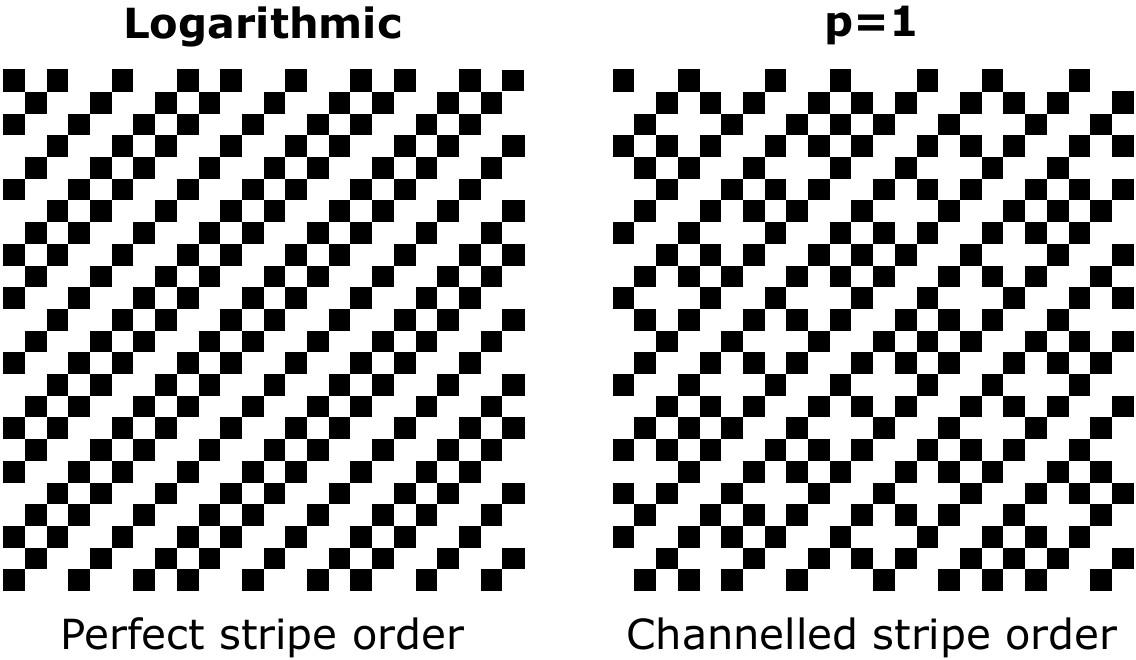}
	\caption{\label{7a} Detail of the charge configurations of the lowest energy states at a density of $n=29/77 \approx 0.377$ and $L=154$. For logarithmic interactions the perfect stripe order is $9 \times 10^{-5} \%$ lower in energy than the channelled stripe order, in contrast to Ref. \cite{Lee:2001p5312} who finds channelled stripes here. For $p=1$ interactions the channelled stripes are $0.0014\%$ lower in energy than the perfect stripes.}
\end{figure}

Here the logarithmic interactions seem to play a special role, in that for a much larger density regime than for algebraic interactions the stripe phases seem to be stable. We find, contradictory to the results of Ref. \cite{Lee:2001p5312}, only stripe order of the Hubbard kind and no channelled stripes in the region between $1/4$ and $4/9$, see Figure \ref{7a}. Only the $1/3$-stripe order seemed to be present for a larger region of densities. We have found `partially filled diagonal channels', or equivalently doped stripe orders, only for interaction types $p=1$ and $p=2$ in a very narrow density range. Our approach differs from Ref. \cite{Lee:2001p5312} in two aspects: we have looked at the ground state, while they looked at low temperature results obtained by simulated annealing only, and we considered non-local particle swaps in the configuration while they restricted the system to local swaps only. We have compared the energy of the ground states we found with the explicit ground states of Ref. \cite{Lee:2001p5312}, taking into account the specific commensurability with the finite lattice size of their doped stripe orders. Figure \ref{7a} gives an example of this energy comparison at $n=\frac{29}{77}$. For all densities that we checked, it was found that our ground state energies are lower - be it only by a very small amount.

Notice that for algebraic interactions, the stripe phase shifts to higher densities when $p$ increases. For large $p$, the stripe phase seems to dominate in the whole region between $1/3$ and $1/2$ fillings.

We also computed the ground state phase diagram for interactions with a finite screening length of the form
\begin{equation}
	V (r) = \frac{e^{- r / \lambda}}{r^p}
\end{equation}
turning into a Bessel function $\mathrm{K}_0(r/\lambda)$ for the `screened logarithmic' interaction. We find that the corresponding phase diagram for screened interactions is largely similar to Figure \ref{PD}, with only small quantitative differences that do grow with decreasing screening length $\lambda$.

\section{Finite temperature}
\label{SectionFiniteT}

The phases described in the previous section survive at finite temperatures, because we are dealing with a system with discrete symmetry in two dimensions. We have computed a finite temperature mean field phase diagram using standard mean field theory\cite{Yeomans}, for the $1/r$ interaction ($p=1$).

It is of course highly questionable whether mean field theory correctly describes the competition between various phases. For example the rich phase diagram of the ANNNI model is constructed using mean field theory\cite{BAK:1982p5293}, but corrections beyond  mean field are shown to tip the delicate balance between different phases and reduce the critical temperature\cite{Desimone:1985p5484}. At the same time Monte Carlo simulations of the ANNNI miss incommensurate phases that are clearly present in analytical extensions of mean field theory\cite{BAK:1982p5293}. Results from the ANNNI model thus suggests that mean field theory acts as a qualitatively reliable first approximation towards the understanding of complex ordering patterns. 


Let us now briefly summarize the quintessence of our mean field theory. For the model Hamiltonian (\ref{LGModel}) we postulate an ansatz for the density,
\begin{equation}
	\langle n_i \rangle = \overline{n} + \sum_{\alpha} m_\alpha \cos (Q_\alpha \cdot r_i)
	\label{Ansatz}
\end{equation}
where there can be as many ordering wave vectors $Q_\alpha$ as one needs to correctly describe the specific charge order. For example, the 1/9 order has ordering wave vectors $Q_1 = \left( 0,\frac{2 \pi}{3} \right)$, $Q_2 = \left( \frac{2 \pi}{3}, \frac{2 \pi}{9}\right)$ and all linear combinations inside the first Brillouin zone. When $m_\alpha \neq 0$ the translational invariance is spontaneously broken. Using the ansatz (\ref{Ansatz}) one constructs a mean field Hamiltonian
\begin{equation}
	H_0 = \sum_i \left[- \mu + 
		\sum_\alpha m_\alpha V_{Q_\alpha} \cos (Q_\alpha \cdot r_i) \right] n_i.
\end{equation}
We then minimise the thermodynamic potential
\begin{equation}
	\Phi = \mathcal{F}_0 + \langle H - H_0 \rangle_0
	\label{ThermoPot}
\end{equation}
with respect to the mean field parameters $m_\alpha$, where $\mathcal{F}_0 = - \frac{1}{\beta} \log \mathrm{Tr} \, e^{-\beta H_0}$ and $\langle \ldots \rangle_0$ implies a thermal average with respect to the mean field Hamiltonian $H_0$. Every charge order we found at zero temperature acts as a possible ansatz, and we numerically minimize at each temperature and density the thermodynamic potential $\Phi$.

Since mean field theory gives only a qualitative phase diagram, and because the zero-temperature phase diagram of Figure \ref{PD} suggests little qualitative difference between various interaction ranges, we restrict ourselves to the Coulomb interaction $V=1/r$. The resulting phase diagram is shown in Figure \ref{PDfiniteT}. In addition to the $1/9$ Wigner crystal phase we also considered $1/12$, $1/16$, $1/20$ and $1/25$ crystals. As for the stripe phases, we only studied the $1/3$ and the $3/7$ `channeled' state. The phase diagram we thus find indeed matches the zero-temperature phase diagram obtained by numerical simulations of the previous section. We emphasize that further studies are needed to understand the possible incommensurate stripe phases in between $1/3$ and $4/7$ filling.

\begin{figure}
	\includegraphics[width=\columnwidth]{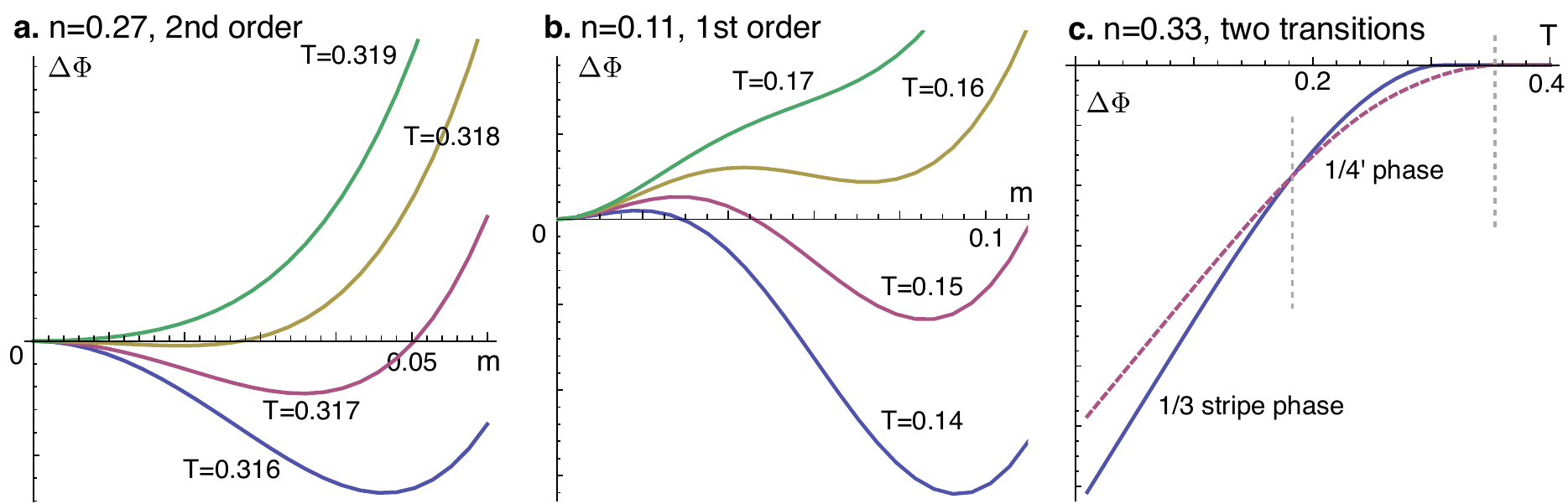}
	\caption{\label{FiniteTEnergy}(Color online) Thermodynamic potential relative to the disordered state $\Delta \Phi$, in arbitrary units, plotted around various phase transitions.
	\textbf{a.} At $n=0.27$ there is a second order phase transition into the $1/4'$ state, clearly visible by the Mexican hat potential. The second order phase transition is common for transitions in the Ising universality class.
	\textbf{b.} At $n=1/9$ there is a first order transition toward the Wigner crystal, as is customary for solidification transitions.
	\textbf{c.} At $n=1/3$ there are two transitions: a second order transition into the 1/4' state followed by a first order transition towards the stripe phase. The 1/4' and the stripe phases are locally stable however for a longer range of temperatures. This implies the possibility of supercooling the 1/4' phase.}
\end{figure}

The transition to the checkerboard phase and the similar 'checkerboard-in-a-checkerboard' $1/4'$ phase is of the second order type within the Ising universality class\cite{Mobius:2009p5314}. The thermodynamic potential for some temperatures around $T_c$, with its typical second order transition behaviour, is shown in Figure \ref{FiniteTEnergy}a. The question arises whether this $1/4'$ `phase' is an artefact of the mean field theory. As is known at half-filling, a liquid-like state with local checkerboard order exists in the presence of long-range interactions\cite{Pramudya:2011p5399}. Such correlated liquid phase can also be present away from half-filling, but will be beyond the scope of standard mean field theory. 

The phase transitions to more complicated orders are always of the first order kind, an example of which is shown in Figure \ref{FiniteTEnergy}b at 1/9 filling. This is a natural result because transitions from solid to liquid phases are usually discontinuous\cite{Brazovskii1974}. The existence of such first order transition implies that one can supercool the high-temperature (or 1/4') state\cite{Das:2011}. At $\rho=1/3$ for example, the $1/4'$ phase remains a local minimum of the free energy even though its energy is higher than that of the stripe phase, see Figure \ref{FiniteTEnergy}c. This suggests that it might be hard to actually trap the system in the lowest energy state.

In combination with our earlier observation that a correlated liquid-like state at intermediate temperatures might exist, we then also expect glassy physics upon supercooling\cite{Schmalian:2000p5453}. With `glassy' we mean that there are macroscopically many local minima of the free energy, leading to slow relaxation rates. There is a difference however with the glassy physics found at $\rho=1/6$ densities\cite{Lee:2002p5305}. There the glassy nature is a ground state property, where glassy physics around a first order transition vanishes if the temperature is low enough.

The mean field theory shows the possibility of supercooling, the consequences thereof such as possible glass-like behavior needs to be addressed differently. Finite temperature numerical simulations however have the great disadvantage that they get easily stuck in such a complicated free energy landscape. It remains thus an open challenge to quantitatively describe the finite temperature phase diagram of the long-range Ising model away from half-filling.

\section{Conclusions and outlook}

In conclusion, we numerically found a ground state and finite temperature phase diagram of the lattice gas model at fixed density on a square lattice with general long-range interactions. We were motivated by the potentiality of nontrivial charge ordering phenomena. Most notable ordering patterns are the generalized Wigner crystals at low densities, supplanted by the stripe order at densities between $1/4$ and $1/2$. All phases are shown in Figure \ref{PD} at zero temperature and in Figure \ref{PDfiniteT} for finite temperatures.

These results extend mainly the work of Ref. \cite{Lee:2001p5312}, in that we have derived complex ordering patterns in the absence of anisotropy or competing interactions. In this case, we suggest that the frustration between the underlying square lattice and the preferable Wigner crystalline state causes the complex ordering. In the vicinity of half-filling this mechanism is supplanted by periodic domain walls in the checkerboard phase. It is these domain walls that cause the formation of stripes.

The finite temperature phase diagram has been obtained using mean field theory, yielding only a qualitative description. Numerical and/or analytical extensions of the classical mean field theory will increase the accuracy of the finite temperature phase diagram. Thereby one can address the possibility of supercooling and glassy physics.

A possible next step is to include the kinetic energy of the particles present. This also allows for an extension to quantum particles\cite{Sengupta:2005p5299} or $O(n)$ spin variables\cite{Nussinov} instead of the classical particles we have considered thusfar.

\section*{Acknowledgements}
L.R. was supported by the Dutch NWO foundation through a VICI grant of Hans Hilgenkamp (University Twente). V.D. and Y.P. were supported by the NSF Grant No. DMR-1005751. The authors wish to thank Henk Bl\"{o}te for fruitful discussions.

\end{document}